\documentclass[
twocolumn,
aps,prd,
nofootinbib,
superscriptaddress,
showpacs,
tightenlines,
amsmath,
amssymb
]{revtex4}
\usepackage{bm}
\usepackage[dvips]{color,graphicx}

\begin{document}
                               

\title{Connection between In Medium Nucleon Form Factors and Deep Inelastic
Structure Functions}

\author{S.~Liuti}
\email[]{sl4y@virginia.edu}

\affiliation{Physics Department, University of Virginia, 382 McCormick Rd., 
Charlottesville, Virginia 22904, USA.}


\begin{abstract}
We present a 
connection between the modifications induced by the nuclear medium 
of the nucleon form factors and of the deep inelastic 
structure functions, 
obtained using the concept of generalized parton distributions. 
Generalized parton distributions allow us to access elements of the partonic 
structure that are common to both the hard inclusive and exclusive scattering
processes in nuclei.
\end{abstract}

\maketitle
An important question that remains currently unsolved in Quantum 
Chromodynamics (QCD), is 
the determination of the quark and
gluon structure of nuclei. 
Theoretical efforts have so far concentrated on two apparently distinct areas.
Many studies were dedicated on one side to pinning down the 
mechanisms producing nuclear medium modifications  
of quark and gluon momentum distributions  
from inclusive (and semi-inclusive) deep inelastic scattering experiments.
Sizable $A$-dependent effects have in fact been observed during the past two decades
(see the review in Ref.\cite{Vogetal}), clearly indicating 
a non-trivial deep inelastic structure of the nucleus 
beyond its naive description as a collection
of weakly bound nucleons whose
quark and gluon structure is unaffected by the nuclear forces.  
On the other side, a number of  
quasi-elastic both inclusive \cite{coul} and 
exclusive \cite{pol} electron-nucleus scattering experiments
allow for investigations of the nucleon form factors of bound nucleons. 
Their initial outcome is also suggestive of non trivial deformations
of the charge and magnetic current distributions of nucleons inside the
nuclear medium. 

\begin{figure}
\includegraphics[width=7.5cm]{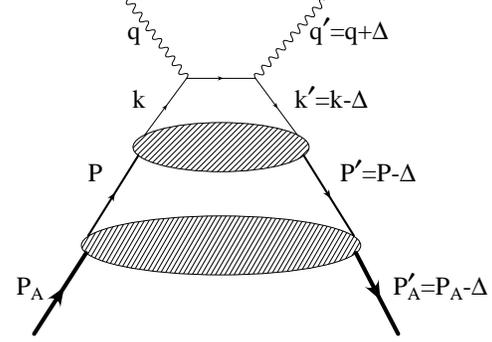}
\caption{Nuclear DVCS amplitude at leading order in $Q^2$. 
For a straightforward presentation of the kinematics 
only the nuclear Impulse Approximation (IA) diagram is shown. 
Modifications to IA are discussed in the text.} 
\label{fig1}
\end{figure}

Recently, a more comprehensive object, the Generalized Parton Distribution (GPD) 
was introduced that interpolates between the Parton Distribution Functions (PDFs) 
from Deep Inelastic Scattering (DIS), and the hadronic form factors 
\cite{GPD}.
A connection was subsequently unraveled \cite{PirRal,Bur1} between GPDs
and the impact parameter dependent parton distributions, 
which allows us to study simultaneously both the light cone momentum and transverse spatial 
distributions of partons.
At leading order, GPDs in nuclei are best visualized 
in terms of the soft parts in the Deeply Virtual 
Compton Scattering (DVCS) process 
described in Fig.\ref{fig1}.

GPDs allow us to make the connection, in this paper, between the in medium 
modifications of the nucleon form factor, and the modifications of the
deep inelastic structure functions of bound nucleons. 
We will consider $^4$He as an ideal nuclear target for our calculations,
since its binding energy per nucleon is strong compared to other few nucleon systems, and at the same time, 
by avoiding extra complications due to non zero spin components
it allows us to focus more directly on nuclear medium modifications.
Evaluations of GPDs in spin zero nuclei were performed in 
\cite{LiuTan2,GuzStr,KirMul}. 
To calculate the amplitude for nuclear DVCS we define (see Fig.\ref{fig1}):  
$P_A$, $P$, $k$, representing the nuclear, the active nucleon's and  
active quark's four-momenta;
$P_A^\prime=P_A-\Delta$, $P^\prime=P-\Delta$, and 
$k^\prime=k-\Delta$ are the final nuclear, nucleon's and quark momenta,
respectively; $q$ is the virtual photon momentum, and $q^\prime=q+\Delta$ is 
the final photon's momentum. 
The relevant invariants are: 
$X  =  k^+/(P_A^+/A)$,  
$Y  =  P^+/(P_A^+/A)$, 
$X_N  =  X/Y \equiv  k^+/P^+$, 
$\zeta  =  - \Delta^+/(P_A^+/A)$, 
$\zeta_N  =  \zeta/Y \equiv  - \Delta^+/P^+$, and 
$q^2 = -Q^2$, $t = -\Delta^2$ \cite{GPD} (we use the notation: $p^\pm = 
1/\sqrt{2}(p_o \pm p_3)$, with
$\displaystyle (pk) = p^+k^- + p^-k^+ - p_\perp \cdot k_\perp$). 
In this paper we focus on the quark region defined in a nucleus by 
$\zeta < X < A$, 
whereby the GPD for a spin zero nucleus reads \cite{LiuTan2}: 
\begin{widetext}
\begin{equation}
H^A(X,\zeta,t)  =  \int \frac{dY d^2 \, {\bf P}_\perp}{2(2\pi)^3}
\frac{{\cal A}}{(A-Y)}  \rho_A 
\left( Y, P^2, t \right) 
\sqrt{\frac{Y-\zeta}{Y}} \left[\hat{H}^N\left(\frac{X}{Y},\frac{\zeta}{Y},P^2,t \right) -
\frac{1}{4} \frac{\left(\zeta/Y \right)^2}{1-\zeta/Y} 
\, \hat{E}^N\left( \frac{X}{Y},\frac{\zeta}{Y},P^2,t \right) \right]. 
\label{HA}
\end{equation}
\end{widetext}
$\rho_A(Y,P^2,t)$ is the light cone (LC) nucleon momentum distribution which
depends explicitly on 
the nucleon's virtuality, $P^2 \equiv P^2(Y,P_\perp^2) \neq M_N^2$; 
${\cal A}$ is a kinematical factor \cite{LiuTan2}. 
$\hat{H}^N$ and $\hat{E}^N$ are the GPDs for the bound nucleon which are intrinsically
modified with respect to the free nucleon ones.
They are assumed to have a flavor decomposition similar to the on-shell case:
\begin{equation}
\hat{H}^N = \frac{2}{3} \hat{H}_u - \frac{1}{3} \hat{H}_d -\frac{1}{3}\hat{H}_s
\label{flavor}
\end{equation}
The nuclear PDFs, $q^A(X)$, and form factor, $F^A(t)$, are 
defined from Eq.(\ref{HA}) by respectively taking the limit
$(\zeta, t) \rightarrow 0$, and by integrating $H^A$ over $X$:
\begin{eqnarray}
q^A(X) & = & H^A_q(X,0) \; \; \; q=u,d,s 
\label{limit1}  
\\
F^A(t) & = & \int\limits_0^A dX H^A(X,t).
\label{limit2}  
\end{eqnarray}   
Since we are interested in the connection with the 
form factor, we consider $\zeta=0$, so that Eq.(\ref{HA}) reduces to:
\begin{equation}
\displaystyle
H^A(X,t) = \int\limits_X^A dY d ^2{\bf P}_\perp \, \rho_A(Y,P^2,t) 
\hat{H}^N\left(\frac{X}{Y},P^2,t \right),
\label{HA-OFF}
\end{equation}
If the intrinsic modifications of the nucleon GPD are disregarded, {\it i.e.}
by assuming $\hat{H}^N(X_N,P^2,t) \approx H^N(X_N,t)$, 
$H_A$ is further reduced to a LC longitudinal convolution:
\begin{eqnarray}
\displaystyle
H^A_{LC}(X,t) & = & \int\limits_X^A dY \, f_A(Y,t) 
H^N\left(\frac{X}{Y},t \right), 
\label{HA-ON}
\\
f_A(Y,t) & = & \int d^2{\bf P}_\perp \rho_A(Y,P^2,t)
\label{fA-ON}
\end{eqnarray}
Deviations from LC convolution are crucial for the description of
both the ``forward'' \cite{AKL,GL92,KPW94} and ``off-forward'' \cite{LiuTan2} 
EMC effects. 
The approach summarized in Eqs.(\ref{HA-ON},\ref{fA-ON}) 
in fact fails to
describe nuclear DIS \cite{TMSG,miller}. 
A number of mean field based calculations \cite{miller,cloet}
have instead been put forth that take into account modifications of
nucleon structure due to the joint effect of attractive scalar and repulsive vector
interactions generated in the average field of the nucleus. 
These impact Eq.(\ref{HA-ON}) through the replacement: 
$H^N(X_N,t) \rightarrow H^N[X_N,t,P^2(\rho)]$, $\rho$ being the nuclear density.
Based on precisely this type of off-shell effect the authors of \cite{miller,cloet}
provide an explanation of both nuclear DIS and 
in medium form factors.  

In our approach, we propose that transverse degrees of freedom, that are 
introduced in a nucleus along with off-shell effects, 
do not decouple as in Eqs.(\ref{HA-ON}) and (\ref{fA-ON})
by being merely integrated over, but 
they play an explicit role.
We illustrate the mechanism for obtaining what we define 
as ``active-$k_\perp$'' effects,
within Hard Scattering Factorization where 
the free nucleon GPD, $H_q$, can be written as: 
\begin{eqnarray}
\displaystyle
H_q & = & \frac{X}{1-X} \int d k_X^2 \int \frac{d ^2 {\bf k}_\perp}{(2 \pi)^3}
\rho_q(k^2,k^{\prime \, 2},k_X^2)
\label{q:spec:on} 
\end{eqnarray}
$k^2$ and $k^{\prime \, 2}$ are the initial and final quarks virtualities related to $k_\perp^2$,
and $({\bf k}_\perp+(1-X) \Delta)^2$, respectively; 
$k_X=P-k$; $\rho_q \propto \mathrm{Tr}\{ \gamma^+ {\cal M} \}$, ${\cal M}$ being
the Fourier transform of the correlator at the nucleon blob.

We assume that nuclear medium modifications of $H_{q}$ originate
similarly to the forward case, with a few important differences that we 
discuss below. 
At large $X$ ($X \gtrsim 0.2$) 
we adopt a quark-diquark model for the in medium $\hat{H}_q$, 
the diquark being a scalar with fixed mass, $k_X^2 \equiv M_X^2$.
The relationship between the quarks virtualities and transverse momenta is shifted 
in a nucleus with respect to the free nucleon one in an $A$-dependent way:
\[
 k^2   =   X_N P^2 - \frac{X_N}{1-X_N} M_X^2 - 
\frac{({\bf k}_\perp - X_N {\bf P}_\perp)^2}{1-X_N}, 
\]
with $P^2=[(Y/A)M_A^2-(M_{A-1}^2+P_\perp^2)Y/(A-Y) - P_\perp^2]$, 
$M_A$ and $M_{A-1}$, being the masses of the initial nucleus and of 
the spectator $A-1$ nucleons system, 
respectively (similar relations hold for $k^{\prime \, 2}$).
This generates a further ``rescaling'' in the $X$ dependence of
$\hat{H}_q$ in a bound nucleon 
that adds on to the relatively small rescaling effect implicit in the LC convolution (\ref{HA-ON}).
As a result, active-$k_\perp$ kinematical effects enhance the so-called binding 
correction to the bound nucleon structure functions.

\begin{figure}
\includegraphics[width=7.cm]{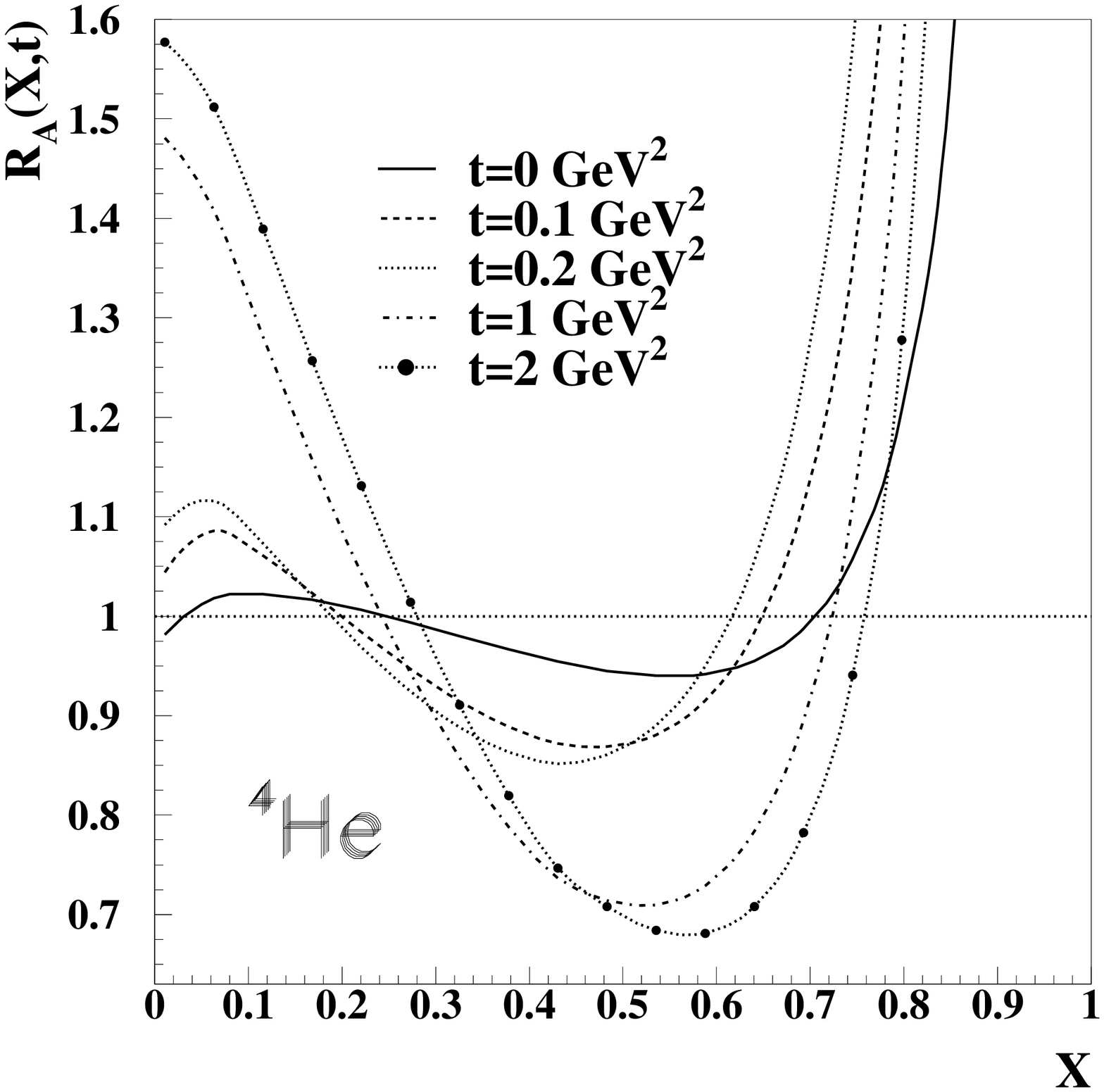}
\includegraphics[width=7.cm]{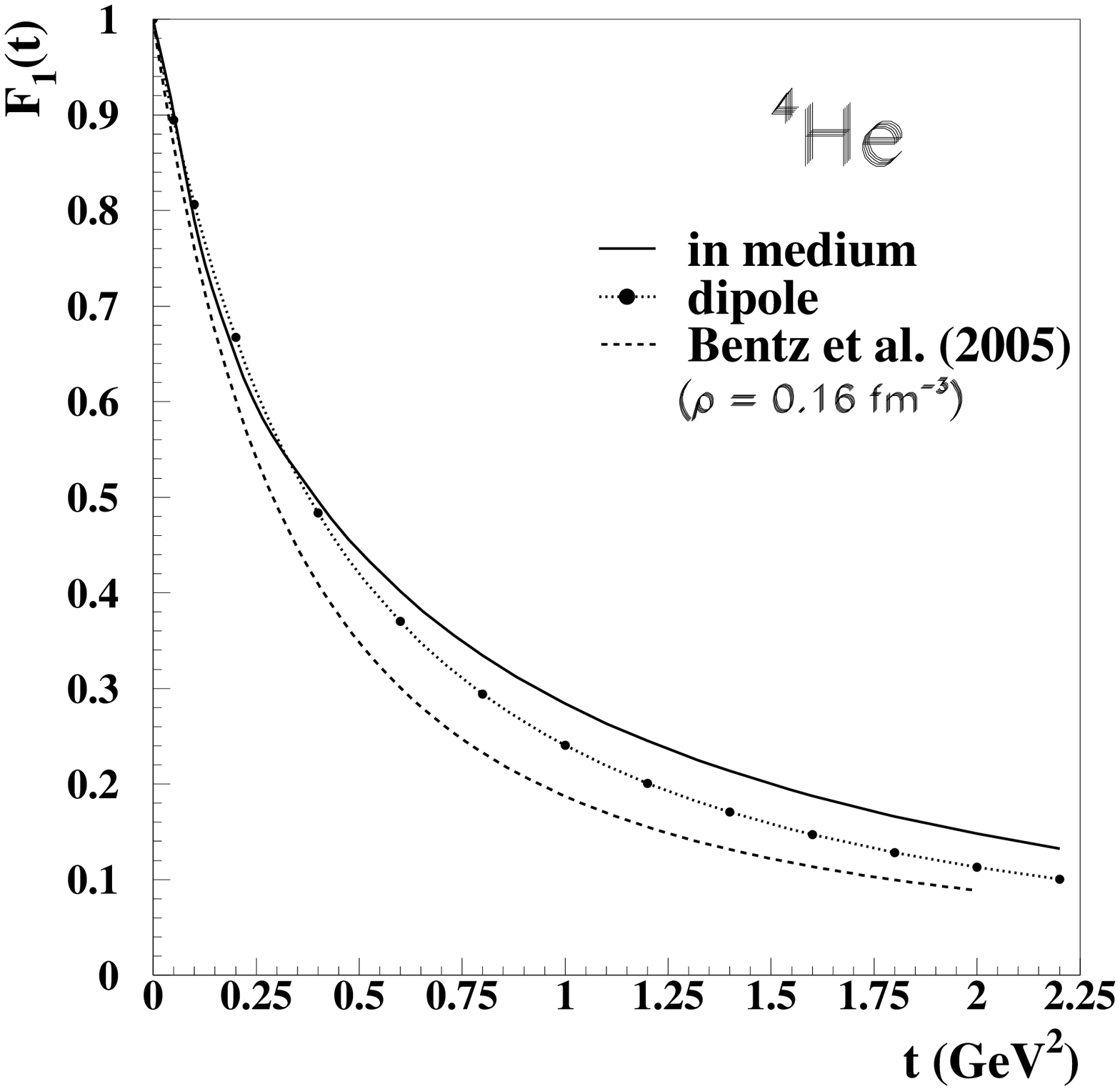}
\caption{
{\bf (a)} Generalized EMC-effect in $^4$He
given by the ratio $R^A$ defined in 
Eq.(\ref{RA}), plotted vs. $X$ for different values of $t$. The 
linear scale in $X$ emphasizes the antishadowing and EMC effect 
parts; 
{\bf (b)} Nucleon form factor in $^4$He. The full curve is our calculation.
The dashed curve is the calculation of Ref.\cite{Bentz} (rescaled as explained in text); a 
similar quenching was  obtained in \cite{miller}.
The dots
correspond to the dipole parametrization of the free nucleon form factor.}
\label{fig2}
\end{figure}

Nuclear modifications are best presented in the off-forward case 
by plotting the ratio of the nuclear GPD over 
the nucleon one, normalized 
by their corresponding form factors: 
\begin{equation}
R_A(X,t) = \frac{H^A(X,t)/F^A(t)}{H^N(X,t)/F^N(t)} 
\label{RA}
\end{equation}
The kinematical effects are visible as a 
dip in the ratios plotted in Fig.\ref{fig2}a.
The increase of the dip with $t$ 
can be 
explained by observing that the off-shellness+binding correction, ${\cal B}_A(t)$:
\begin{equation}
{\cal B}_A(t) = \frac{\int_0^A dY \, \rho_A(Y,\langle P^2 \rangle ,t) (1-Y)}{\int_0^A dY \, \rho_A(Y,\langle P^2 \rangle ,t)}
\label{BA}
\end{equation}
grows (in our model almost linearly) with $t$.
When active-$k_\perp$ effects are considered, {\it i.e.} 
$\langle P^2 \rangle \neq M_N^2$, one obtains a bigger 
shift in $X$, and therefore an enhanced effect displayed in Fig.\ref{fig2}a. 

Intrinsic dynamical modifications allow us in principle, to access
the nucleon's spatial deformations inside the nuclear medium.
The latter have been so far taken into account in a variety of models 
by shifting the parameters that either represent directly, or are
related to the confinement size in hadrons \cite{Pirner}. 
GPDs because of their interpretation in coordinate space, 
supersede {\em ad hoc} models by allowing for 
a direct evaluation of the in medium spatial distributions of hadrons
\cite{NUC05}.  
In the approach presented here, the origin 
of dynamical modifications is associated with  
the $k_\perp$ dependent parton re-interactions generating 
shadowing (suppression of the in medium structure function
with respect to the free nucleon one) and antishadowing (enhancement), 
at low $X$ ($X \lesssim 0.2$).  
By inspecting the energy denominators 
dominating the lepton-nucleus scattering process, one sees that the  
mass of the spectator component is in this case, $k_X^2 \equiv s \propto 1/X$ \cite{DokKho}.  
$\rho_q(k^2,k^{\prime \, 2},k_X^2)$ in Eq.(\ref{q:spec:on}) is calculated
in a Regge theory based model for the 
quark-nucleon amplitude, ${\cal M} \approx T_{qN}$
\cite{BroSch,NNN}. 
In forward lepton-nucleus scattering
$T_{qN}$ is, in fact, assumed to have similar analytical properties as 
the proton-proton (proton-antiproton) 
amplitude. One can express it therefore in terms of Pomeron (P), 
Odderon (O), and other Reggeon (R) exchanges. As shown in \cite{NNN,BroSch}
nuclear shadowing and antishadowing arise because at low $X$ 
the struck partonic configuration
lives over several intra-nuclear distances and,  
as a consequence, it undergoes multiple scattering.
The nuclear amplitude $T_{qA}$ reflects
a rather complicated combination of 
destructive (shadowing) and constructive (antishadowing) contributions  
from the different exchange terms  
entering Glauber's multiple scattering series with same or opposite phases, 
respectively.
In this paper we model the nuclear GPDs at low $X$ and $\zeta=0$ 
by directly extrapolating from Ref.\cite{BroSch}, 
{\it i.e.} by replacing the forward $qN$ amplitudes used in \cite{BroSch} 
with the $t \neq 0$ ones.
Because of the absence of skewedness, we expect in fact 
the longitudinal distances that constitute the driving mechanism behind
shadowing/antishadowing models, to behave similarly to the forward case.  
As for the $t$ dependence, we obtain an increase
with $t$ in the antishadowing that tends to balance 
the effect at large $X$ (Fig.\ref{fig2}a), 
or a continuation of the trend set by fixing
the parameters for $t=0$. 
The interplay between kinematical and dynamical effects might lead  
to a dominance of either the large $X$ part (depletion)
or the low $X$ part (enhancement) of the structure function,
lacking, in this case the constraint from baryon number 
conservation. 
Finally, we point out the difference between our approach and the one in Ref.\cite{FreStr}
where nuclear modifications were estimated in a completely different situation, namely at 
$t=0$, and $\zeta \neq 0$.

The modifications thus obtained by
explicitly considering active-$k_\perp$ effects 
impact directly the in medium form
factor. We define it as:
\begin{equation}
\hat{F}^N_1(t) = \left[ \frac{F^A(t)}{F^A_{LC}(t)} \right] F^N_1(t)
\label{inmedium}
\end{equation}
where:
\begin{eqnarray}
F^A_{LC}(t)  =  \int\limits_0^A H^A_{LC}(X,t) dX = F^N_1(t) \int\limits_0^A f_A(Y,t) dY .
\label{FF-LC}
\end{eqnarray}
$F^A(t)$ in Eq.(\ref{inmedium}) is given by Eqs.(\ref{limit2}) and (\ref{HA-OFF}); 
$F^A_{LC}$ is calculated using Eq.(\ref{HA-ON}); 
$F^N_1$ is the free nucleon form factor. 
According to our description of the $t$ dependence of GPDs (Fig.\ref{fig2}a),
at $t \approx 0.5$, the antishadowing enhancement  
starts dominating over the depletion given by both shadowing and the EMC effect, 
resulting in larger values for the in medium form factor 
($\hat{F}_1^N > F_1^N$, Fig.\ref{fig2}b). The dominance of antishadowing 
can also be understood by analyzing the $X$ values that govern the form factor. 
As shown in \cite{LiuTan1,DieKro}, the average value, $\langle X(t) \rangle$, 
increases with $t$ with a slope such that at $t \lesssim 4$ GeV$^2$ the antishadowing
region ($0.02 <X < 0.2$) is dominating. 
In Fig.\ref{fig2}b we show a comparison with the results obtained in Ref.\cite{Bentz}
using a Nambu-Jona-Lasinio (NJL) based effective quark theory (notice that we appropriately 
rescaled the curves shown in \cite{Bentz} 
for a consistent comparison). Although there is
agreement between the two models at low $t$, a discrepancy occurs around $t = 0.5$,
perhaps due to the usage of the quark-diquark model in \cite{Bentz}, 
as opposed to the shadowing/antishadowing description in our calculation.   

We are, therefore, able to unravel a common origin between the modifications of the 
deep inelastic structure and the form factor of the proton. In our
interpretation a key role is played by $k_\perp$-dependent, leading order 
parton re-interactions in a nucleus, governed by Regge-type exchanges, at variance
with the meson exchanges of Refs.\cite{miller,cloet,Bentz}. 
Our findings point to a form of duality 
perhaps in a similar direction 
as recent phenomenological studies performed in \cite{Polya}.

What are the consequences of our results for phenomenology? 
In medium effects have been investigated in quasi-elastic electron scattering,
both through the measurement of the Coulomb Sum Rule \cite{coul}, 
and in polarization transfer experiments yielding the ratio of 
the nucleon's electric and magnetic form factors through its proportionality to the  
transverse to longitudinal transfer polarizations \cite{pol}. 
Despite definitive interpretations of the inclusive data 
are, so far, pending on the treatment of Coulomb corrections \cite{Zein},
there seems to be an indication of a suppression of the form factor  
for $t \equiv Q^2 \lesssim 0.3$ GeV$^2$. Similarly, 
the polarization transfer experiments
also show a depletion at low $t$ that seems to however saturate at larger $t$.
Our results, concentrating on the nucleon's Dirac form factor, 
are so far consistent with these findings. We furthermore  
predict the onset of saturation and an enhancement 
at larger values of $t$ of interest for future proposals \cite{Chen}.
 

In conclusion, we presented a connection using GPDs, 
between the in medium nucleon form factors, and 
the nuclear deep inelastic structure functions. 
The in medium form factor deformation
follows a pattern in four-momentum transfer, $t$,
with an initial suppression, up to $t \approx 0.5$, 
and a subsequent enhancement. 
This pattern is a direct consequence of the behavior 
with $X$ of the zero skewedness 
GPD, $H^A(X,t)$ which, because of the coupling in our model to 
transverse degrees of freedom in a nucleus ($k_\perp$), 
displays enhanced antishadowing and EMC effects with increasing $t$.   
This reflects on the form factor's behavior that, being given by 
the integral of the nuclear GPD, results from a balance of 
the same physical effects
which are known to be present in deep inelastic scattering from 
nuclei. 

I am indebted to Z.E. Meziani, N. Nikolaev, J. Ralston  and A. Thomas for  
critical comments and discussions. I also thank W. Bentz for providing 
me with his calculations of the in medium form factors.
This work is supported by the U.S. Department
of Energy grant no. DE-FG02-01ER41200. 

\end{document}